\documentclass[epjST]{svjour}
\usepackage{graphicx}
\usepackage{amsmath}
\usepackage{mathtools}
\usepackage{amssymb}
\usepackage{bm}
\usepackage{pifont}
\usepackage{color}
\usepackage[FIGTOPCAP]{subfigure}
\usepackage[latin1]{inputenc}

\newcommand{\omi}{\mbox{\boldmath $\Omega_i$}}

\newcommand{\HT}{\mbox{\boldmath $H$}}
\newcommand{\QTa}{\mbox{\boldmath $Q_1$}}
\newcommand{\QTb}{\mbox{\boldmath $Q_2$}}

\begin{document}

\title{Large density expansion of a hydrodynamic theory for self-propelled particles}

\author{Thomas Ihle\inst{1}\fnmsep\thanks{\email{thomas.ihle@ndsu.edu}}}
\institute{Department of Physics, North Dakota State University, Fargo, ND 58108-6050, USA}
\abstract{
Recently, an Enskog-type kinetic theory for Vicsek-type models for self-propelled particles has been proposed 
[T. Ihle, Phys. Rev. E {\bf 83}, 030901 (2011)].
This theory is based on an exact equation for a Markov chain in phase space and 
is not limited to small density.
Previously, the hydrodynamic equations were derived from this theory and its transport coefficients were given in terms
of infinite series. Here, I show that the transport coefficients take a simple form in the large density limit.
This allows me to analytically evaluate the well-known density instability of the polarly ordered phase near the flocking threshold 
at moderate and large densities.
The growth rate of a longitudinal perturbation is calculated and several scaling regimes, including three different power laws, are identified.
It is shown that at large densities, the restabilization of the ordered phase at smaller noise is analytically accessible within the range of 
validity of the hydrodynamic theory.
Analytical predictions for the width of the unstable band, the maximum growth rate and for the wave number below which the instability occurs are given.
In particular, the system size below which spatial perturbations of the homogeneous ordered state are stable is predicted to scale with $\sqrt{M}$
where
$M$ is the average number of collision partners. The typical 
time scale until the instability becomes visible is calculated and is proportional to $M$.
}

\maketitle

\section{Introduction}
\label{intro}

The emergence of collective motion of living things, such as insects, birds, slime molds, bacteria
and fish is a fascinating far-from-equilibrium phenomenon which has attracted a great deal of cross-disciplinary attention 
\cite{vicsek_zafeiris_12}.
The study of these systems falls under the broader umbrella of ``active matter''.
This term stands for collections of agents that are able to extract and 
dissipate energy from their surrounding to produce systematic motion \cite{ramaswamy_10,marchetti_13}.
Non-living examples of active matter are chemically powered nanorods \cite{rueckner_07}, 
networks of actin fibers driven by molecular 
motors \cite{humphrey_02} and swarms of
interacting robots \cite{rubenstein_14,wilson_14}.
In 1995, a minimal computational 
model that captures the essentials of collective motion without entering too much detail was introduced by Vicsek 
\cite{vicsek_95_97}.
In this so-called Vicsek model (VM), agents are represented as point particles traveling at constant speed. 
The agents follow
noisy interaction rules which
aim to align the velocity vector of any given particle with its neighbors.
Later, more sophisticated models that include additional interactions such as attraction and short-range repulsion
were introduced \cite{levine_00,couzin_02,romenskyy_13,grossmann_13}.
On the theoretical side, coarse-grained descriptions of the dynamics in terms of phenomenological hydrodynamic equations have been proposed on the basis of symmetry and conservation law arguments.Well-knowns examples are the Toner-Tu theory
for polar active matter \cite{toner_98,toner_12} and the theory by Kruse {\em et al.} for active polar gels \cite{kruse_05}.
While being very successful, a disadvantage of these approaches is
that no link between microscopic collision rules and the coefficients of the macroscopic equations is provided. 
In particular, since all coefficients must be related to only a few microscopic parameters, they
cannot be varied independently and the actual parameter space should be more restricted 
than the hydrodynamic equations
might suggest.
Furthermore, by just postulating equations there is the immanent danger that some terms or even entire equations 
might have been ommitted,
which could become relevant in new, previously untested, situations. 
To address the missing-link issue, several groups have derived hydrodynamic equations from the underlying microscopic rules and
provided expressions for the transport coefficients in terms of microscopic parameters 
\cite{aranson_05,bertin_06,mishra_10,ihle_11,menzel_12,romanczuk_12,peshkov_12}.
One of the first attempts to put the Toner-Tu theory
on a microscopic basis was published by Bertin {\em et al.}, who studied a Vicsek-type model with continuous time-dynamics
and binary interactions
by means of a Boltzmann approach \cite{bertin_06,bertin_09}.
While the authors recently clarified \cite{bertin_14A} that even at low density 
their underlying microscopic model is not identical
to the Vicsek-model, many predictions agree qualitatively with the ones for the VM. 
This Boltzmann approach was later extended to systems of self-propelled particles 
with nematic and metric-free interactions, and hydrodynamic equations were drived \cite{peshkov_12,ngo_12}.
Very recently, the very basis of these derivations -- the Boltzmann equation and its validity in active matter --
has been critically assessed \cite{thueroff_13,ihle_14,ihle_14A}.
In particular, it has been shown that, at least near the threshold to collective motion, 
the binary collision assumption is not valid in the VM 
at realistic particle speeds and even at very low densities \cite{ihle_14}.
Furthermore, it has been 
demonstrated that the mean-field assumption of molecular chaos is not justified near the threshold to collective motion 
in soft active colloids
and that the Boltzmann theory must be amended by pre-collisional correlations \cite{hanke_13}.
A first attempt to rigorously include correlations by means of a ring-kinetic theory for Vicsek-type models was put forward 
by Chou {\em et al.} \cite{chou_14}.
This theory is very complicated; work is still in progress to simplify it \cite{ihle_15}.

Nevertheless, arguments from ring-kinetic theory can be used to confirm the plausible presumption that 
mean-field theories become reliable in the VM at large particle speeds (or time steps) and/or at large 
particle densities \cite{FOOT2}. 
Here, large density means that the average number $M$ of collision partners of a given agent
is much larger than one. 
It is known from equilibrium 
statistical mechanics that the behavior of spin-models become mean-field or more mean-field-like 
if the number of interaction partners is increased which can either be achieved by extending the range of interaction or
by increasing the spatial dimension.
It appears reasonable to assume a similar tendency in the VM which consists of moving ``spins''.

The large density limit $M\gg 1$ is beyond the capability of Boltzmann approaches because 
Boltzmann equations are restricted to binary interactions.
However, a recently proposed Enskog-like theory \cite{ihle_11,ihle_13,ihle_14} 
has no limitation on density and has already be applied to a model
with $M=2\ldots 7$ and metric-free interactions \cite{chou_12}.
This Enskog-like theory is based on an exact equation for a Markov chain in phase space.
In a previous paper \cite{ihle_11}, 
hydrodynamic equations were derived from this theory and its transport coefficients were given in terms
of infinite series. 
In this paper, I analyze the transport coefficients in the large density limit, $M\gg 1$, and show 
that they take a simple form.
This allows me to analytically evaluate the well-known density instability of the 
polarly ordered phase near the flocking threshold \cite{chate_04_08,bertin_06}, and to obtain simple formulas and scaling laws for
the dispersion relation.
Note that a similar analysis for the opposite limit $M\ll 1$ has already been performed by Bertin {\em et al.} 
\cite{bertin_09}.

The large density expansion performed in this paper is supposed to be benefitial 
for several reasons.
First, one does not have to worry about the validity of the underlying mean-field assumption of Molecular Chaos:
As discussed above, this assumption is expected to be asymptotically valid for $M\rightarrow \infty$.
Thus, the obtained relations should be quantitatively correct in this limit.
As will be shown below, at large density, the restabilization of the homogeneous ordered phase happens closer 
to the threshold. Hence, this effect 
occurs within the validity domain of the hydrodynamic theory, 
and, aiming at quantitative agreement, one is not forced to evaluate the kinetic eqution directly as proposed 
in Ref. \cite{bertin_09}.
Second, biologically realistic models of fish schools \cite{tegeder_95} and bird flocks \cite{ballerini_08} assume that 
$M$ is between 2 and 7, thus an approach for large $M$ seems promising.
Third, similar to the opposite limit, $M\rightarrow 0$, the large density approach 
leads to a simplification of the expressions for the transport coefficients and allows a physical interpretation of the 
instability close to the threshold of collective motion.

An obvious concern about the large density limit is how realistic it is for active matter in general.
Of course, for example, 
for systems of active colloids with short-ranged excluded volume interactions
it will be problematic. 
However, it does make sense for systems with long-ranged alignment interactions.
Speaking in terms of the terminology put forward by Couzin et al. \cite{couzin_02}, the condition $M\gg 1$ can occur in a model
with a large ``zone of orientations'' and a very small ``zone of repulsion''
Furthermore, in the process of doing the actual calculations and performing expansions in powers of $1/M$, 
one realizes that error terms are often proportional to ${\rm exp}(-M)$.
For example, for $M=2$, this is already a reasonably small number. 
Therefore, it is possible that the large density formulas derived in this paper
could still be valid at $M$ slightly larger than one without loosing too much accuracy.

To the best of my knowledge, almost no analytical results exist for the Vicsek-model at high particle density.
This paper is intended to fill this void. One of the main results 
is the prediction that the homogeneous ordered phase remains stable to small perturbations if the linear system size is chosen to be
smaller than $L^{*}=\pi\sqrt{2M}$. 
Another main result is that in systems larger than $L^{*}$  the number of time steps to develop an instability is 
equal to or larger than $16\pi\,M$.  

\section{Vicsek model}
\label{sec:Vicsek}

In the  two-dimensional Vicsek-model (VM) \cite{vicsek_95_97,nagy_07}, $N$ particles with average number density 
$\rho_0=N/A$ move at constant
speed $v_0$.
The system of area $A$
evolves in discrete time intervals of size $\tau$.
The particles are assumed to have zero volume.
The evolution takes place in of two steps: streaming and collision.
During streaming, the particle positions ${\bf x}_j(t)$
are updated in parallel according to
\begin{equation}
\label{STREAM}
{\bf x}_j(t+\tau)={\bf x}_j(t)+\tau {\bf v}_j(t)\,.
\end{equation}
Because the particle speeds remain the same at all times, the velocities, ${\bf v}_j(t)$, are parametrized by the ``flying'' angles, 
$\theta_j$,
${\bf v}_j=v_0(\cos{\theta_j},\sin{\theta_j})$.
In the collision step,
the directions $\theta_j$ 
are modified in the following way:
a circle of radius $R$ is drawn around the focal particle $j$, and the average direction $\Phi_j$ of motion
of the
particles
within the circle is determined
according to
\begin{equation}
\label{COLLIS}
\Phi_j={\rm Arg}\left\{ 
\sum_{k}{\rm e}^{i\theta_k}\right\}
\end{equation}
where the sum goes over all particles inside the corresponding circle  (including particle $j$). 
The new flying angles 
are then given as,
\begin{equation}
\label{VM_RULE}
\theta_j(t+\tau)=\Phi_j+\xi_j
\end{equation}
 where $\xi_j$ is a random number uniformly distributed in
the interval $[-\eta/2,\eta/2]$. 
In this paper, a forward-upating rule \cite{baglietto_08_09}, corresponding to the so-called standard Vicsek-model, is used:
the updated positions ${\bf x}_j(t+\tau)$ instead of the ``old'' positions ${\bf x}_j(t)$ 
enter the calculations of the average directions $\Phi_j$.
Important dimensionless parameters of the model are (i) the noise strength $\eta$,
(ii) the average number of particles in a collision circle, $M=\pi R^2 \rho_0$, which is basically a normalized density, and (iii)  
the ratio $\gamma=R/\lambda$ of the range $R$ of the interaction to the mean displacement during streaming $\lambda=v_0\tau$ 
which can be interpreted as a mean free path (mfp).

\section{Hydrodynamic theory}
\label{sec:hydro}

\subsection{Macroscopic equations for the Vicsek model}
\label{sec:background}

Using Boltzmann's assumption of molecular chaos, 
an Enskog-type kinetic equation for the one-particle density $f(\theta,{\bf x},t)$
was derived from the exact evolution equation of the $N$-particle probability. 
This was done for the standard Vicsek-model (VM) \cite{ihle_11} as well as for other 
microscopic models \cite{ihle_09,chou_12,romensky_14}.
By means of a Chapman-Enskog expansion and an angular Fourier transformation of the one-particle density,
\begin{equation}
f(\theta,{\bf x},t)=g_0({\bf x},t)+\sum_{k=1}^{\infty}
\left[g_k({\bf x},t)\cos{(k\theta)}+
h_k({\bf x},t)\sin{(k\theta)}\right]\,
\end{equation}
hydrodynamic
equations were obtained for the dimensionless particle density $\rho$ and momentum density $\vec{w}$ of the VM \cite{ihle_11}.
In this formulation, all times are scaled by the time step $\tau$, all lengths are scaled by the mean free path (mfp)
, $\lambda=v_0 \tau$, and  all other quantities are scaled accordingly.
For example, the actual two-dimensional number density $\tilde{\rho}$ and momentum density $\tilde{\bf w}$ can be obtained from their scaled versions as
$\tilde{\rho}=\rho/\lambda^2$ and $\tilde{\bf w}={\bf w} v_0/\lambda^2$.
To emphasize rotational invariance, the hydrodynamic equations are given in terms of the tensors
$\HT$, $\QTa$, and $\QTb$
which depend on spatial derivatives of $\rho$ and $\vec{w}$:
\begin{eqnarray}
\label{CONTIN}
\partial_t\rho+\nabla\cdot {\bf w}
&=&0 \\
\label{NAVIER}
\partial_{t}\vec{w}+\nabla\cdot\HT&=&-b\,\nabla\rho+(\Lambda-1)\vec{w}+
\QTa\cdot \vec{w}+\QTb\cdot\nabla\rho
\end{eqnarray}
with $b=(3-\Lambda)/4$. Here, $\Lambda$ is the amplification factor for the first Fourier mode, which also determines the
linear growth rate $(1-\Lambda)$
of the momentum density. 
This is because density and momentum density are the zeroth and first order moments of the one-particle distribution function, respectively, and
therefore, the components of the momentum density are proportional to the first Fourier coefficients, $w_x\propto g_1$, $w_y\propto h_1$.
In Ref. \cite{ihle_11}, in the thermodynamic limit $N\rightarrow \infty$,
an infinite series was found for $\Lambda$,
\begin{eqnarray}
\nonumber
\Lambda&=&{4 \over \eta}{\rm sin}\left({\eta\over 2}\right)
{\rm e}^{-M_R} \sum_{n=1}^{\infty} {n^2 M_R^{n-1}\over n!}\,I(n) \\
\label{LAMBDA_DEF}
I(n)&=&{1\over (2\pi)^n}
\int_0^{2\pi}d\theta_1\ldots
\int_0^{2\pi}d\theta_n
\,\cos{\bar{\theta}}\,
\cos{\theta_1}
\end{eqnarray}
where $\bar{\theta}=\bar{\theta}(\theta_1,\theta_2,\ldots,\theta_n)$ is equal to the average angle $\Phi_1$ 
defined in Eq. (\ref{COLLIS}).
Here, $M_R$ is the {\em local} average number of particles within a collision circle centered around the position ${\bf x}$, 
\begin{equation}
\label{M_R_DEF}
M_R({\bf x},t)=\int_{|{\bf x}-\bf{x'}|\leq R} \,d{\bf x'} \int_0^{2 \pi} d\theta\, 
f(\theta,{\bf x'},t)
=\int_{|{\bf x}-\bf{x'}|\leq R} \,d{\bf x'} \rho({\bf x'},t)
\end{equation}
For a homogeneous system, one has $M_R=M=\pi R^2 \tilde{\rho}_0$.
If $\Lambda$ is larger than one, the disordered state of zero average motion becomes unstable 
and a polarly ordered state with non-zero global momentum develops.
Thus, setting $\Lambda=1$ defines the threshold noise $\eta_C(M)$ of the flocking transition, at the level of homogeneous mean field theory.
The hydrodynamic equations, Eqs. (\ref{CONTIN}) and (\ref{NAVIER}), 
are only valid in the vicinity of the transition to polar order, that is, for $|\lambda-1|\ll 1$.
This condition was essential to obtain a closed set of just two equations, see Refs. \cite{ihle_11,ihle_14}.
Physically, this assumption means that the first order Fourier coefficients $g_1$ and $h_1$ are evolving so slowly that 
the higher order coefficients, $g_2,g_3,\ldots$, become 
enslaved to them \cite{FOOT1}. 
Therefore, no additional equations for the temporal evolution of the higher order coefficients are needed.

The momentum flux tensor $\HT$ and the tensors $\QTa$, $\QTb$,
\begin{equation}
\label{TENSOR1}
\HT=\sum_{i=1}^5h_i\,\omi\;\;\;\;\;\;\;
\QTa=\sum_{i=1}^5q_i\,\omi \;\;\;\;\;\; 
\QTb=\sum_{i=1}^5k_i\,\omi
\end{equation}
are given in terms of five symmetric traceless tensors $\omi$, 
\begin{eqnarray}
\nonumber
\Omega_{1,\alpha\beta}&=&\partial_{\alpha}w_{\beta}+\partial_{\beta}w_{\alpha}
-\delta_{\alpha\beta}\partial_{\gamma}w_{\gamma} \\
\nonumber
\Omega_{2,\alpha\beta}&=&2\partial_{\alpha}\partial_{\beta}\rho
-\delta_{\alpha\beta}\partial^2_{\gamma}\rho \\
\nonumber
\Omega_{3,\alpha\beta}&=&2w_{\alpha}w_{\beta}
-\delta_{\alpha\beta}w^2 \\ 
\nonumber
\Omega_{4,\alpha\beta}&=&w_{\alpha}\partial_{\beta}\rho+w_{\beta}\partial_{\alpha}\rho
-\delta_{\alpha\beta}w_{\gamma}\partial_{\gamma}\rho \\
\label{OMEGA_DEF}
\Omega_{5,\alpha\beta}&=&2(\partial_{\alpha}\rho)(\partial_{\beta}\rho)
-\delta_{\alpha\beta}(\partial_{\gamma}\rho)^2\,.
\end{eqnarray}
The tensor $\Omega_1$ is the viscous stress tensor of a two-dimensional fluid.
The transport coefficients in Eq.\ (\ref{TENSOR1}) were obtained in the additional limit of large mean free path, $\lambda=v_0\tau\gg R$, and 
are given in 
Table \ref{TAB2}. 
This means, contributions  
from the so-called collisional momentum transfer, which are supposed to be relevant at small mean free paths, see Ref. \cite{ihle_09}, 
are neglected here.
The rationale for this choice was that at small mean free paths and not too large density, the mean-field assumption 
(on which our theory is based) is supposed to break down anyway.

The transport coefficients depend on the following variables,
\begin{eqnarray}
\nonumber
p&=&
{4\over \eta} \sin{(\eta)}\sum_{n=1}^{\infty}
{{\rm e}^{-M_R}\over n!}n^2
M_R^{n-1} J_1(n) \\ 
\nonumber
q&=&
{4\pi \gamma^2\over \eta} \sin{(\eta)}\sum_{n=2}^{\infty}
{{\rm e}^{-M_R}\over n!}n^2(n-1)
M_R^{n-2} J_2(n)\\ 
\label{DEF_PQ}
S&=&
{8\pi\gamma^2\over \eta }\sin{\eta\over 2}
\sum_{n=2}^{\infty}
{{\rm e}^{-M_R}\over n!}
n^2(n-1) M_R^{n-2} J_3(n) \\
\nonumber
\Gamma&=&
{8\pi^2\gamma^4\over 3 \eta}\sin{\eta\over 2}
\sum_{n=3}^{\infty}
{{\rm e}^{-M_R}\over n!}
n^2 (n-1)(n-2)M_R^{n-3} J_4(n)
\end{eqnarray}
\begin{table}
\begin{center}
{\renewcommand{\arraystretch}{1.4}
\large
\begin{tabular}{|r||c|c|c|}
\hline
$j$ & $h_j$      &    $q_j$      &      $k_j$     \\
\hline
\hline
$1$ & ${1+p\over 8(p-1)}$          &  ${S\over 2(p-1)}$     &  $ {S\over 8 (p-1)}$ \\ 
\hline
$2$ & $-{p^2+10p+1\over 96(p-1)^2}$& $-{S\over 4(p-1)^2}$   & $-{S (p+5)\over 96(p-1)^2}$ \\
\hline
$3$ & $-{q\over 2(p-1)} $          & $\Gamma-{Sq\over p-1}$ & ${\Gamma\over 4}-{Sq\over 4(p-1)}$ \\
\hline  
$4$ & ${q(1+p)\over 4(p-1)^2}$     & ${\Gamma\over 2}-{Sq(p-3)\over 2(p-1)^2}$ & ${\Gamma\over 12}-{Sq(p-4)\over 12(p-1)^2}$ \\
\hline
$5$ & $-{q(p^2+10p+1)\over 48(p-1)^3}$ & ${\Gamma\over 24}-{Sq(p^2-2p+13)\over24(p-1)^3}$ & $-{Sq(p+5)\over 48(p-1)^3}$ \\
\hline
\end{tabular}
}
\caption{The transport coefficients $h_j$, $q_j$ and $k_j$,
defined in Eq. (\ref{TENSOR1}), are expressed as functions of 
$\Gamma$, $S$, $p$, $q$, see Eq. (\ref{DEF_PQ}).
}
\label{TAB2}
\end{center}
\vspace*{-5ex}
\end{table}
where $\gamma$ is the ratio of the interaction radius to the mfp, $\gamma=R/\tau v_0$.
Note that these coefficients have a non-local dependence on position through the normalized density $M_R$.
The transport coefficients contain the following four types of $n-$dimensional integrals,
\begin{equation}
\label{INTEGRALS}
J_m(n)={1\over (2\pi)^n}
\int_0^{2\pi}d\theta_1\ldots
\int_0^{2\pi}d\theta_n\,
\Psi_m
\end{equation}
where $\Psi_m$ is given
by $\Psi_1=
\cos^2{\bar{\theta}}\,
\cos{2\theta_1}$,
$\Psi_2=
\cos{\bar{\theta}}\,\sin{\bar{\theta}}\,
\cos{\theta_1}\,\sin{\theta_2}
$,\\
$\Psi_3=
\cos{\bar{\theta}}\,
\cos{\theta_1}\,\cos{2 \theta_2}
$, and
$\Psi_4=
\cos{\bar{\theta}}\,
\cos{\theta_1}\,\cos{\theta_2}\,\cos{\theta_3}$.
The average angle
$\bar{\theta}\equiv \Phi_1$ is a function of the angles
$\theta_1,\theta_2,\ldots \theta_n$, see Eq. (\ref{COLLIS}).

The Navier-Stokes-type equation, Eq.\ (\ref{NAVIER}) is consistent with the one from Toner-Tu theory  
\cite{toner_98,toner_12} but
contains 
additional gradient terms. 
The additional terms are a result of a Chapman-Enskog expansion that includes all terms up to {\em third order} in a formal expansion parameter. 
Discussions of higher order expansions can be found in Refs. \cite{ihle_14,peshkov_14,ihle_14A,bertin_14A}.
Eq. (\ref{NAVIER}) has a simple homogeneous flocking solution: 
$\vec{w}=w_0\,{\bf \hat{n}}$ and $\rho=\rho_0$. The amplitude of the flow is given by 
\begin{equation}
\label{W0_DEF}
w_0=\sqrt{1-\Lambda\over q_3}\,. 
\end{equation}

\subsection{Large $M$ expansion}
\label{sec:largeM}

Using an analogy to the freely-jointed chain model for polymers, 
the angular integrals, Eqs. (\ref{INTEGRALS}), can be asymptotially evaluated in the limit $n\gg 1$ \cite{ihle_15}. 
To leading order, the results are \cite{ihle_10_V1},
\begin{eqnarray}
\nonumber
J_1(n)&\sim &{1\over 8 n} \\
\nonumber
J_2(n)& \sim &{1\over 8 n} \\
\nonumber
J_3(n)& \sim &- {\sqrt{\pi}\over 32 n^{3/2}} \\
\label{J_DEFS}
J_4(n)& \sim &- {3 \sqrt{\pi}\over 32 n^{3/2}} 
\end{eqnarray}
The angular integral in the definition of $\Lambda$, Eq. (\ref{LAMBDA_DEF}), is evaluated similarly,
\begin{equation}
\label{I_ASYM}
I(n)\sim \sqrt{\pi \over 16 n}.
\end{equation}
Even with these approximations, 
the infinite series in Eqs. (\ref{LAMBDA_DEF}) and (\ref{DEF_PQ}), still cannot be calculated exactly. 
However, we note that they can be written as an average over a 
Poisson distribution.
This is a consequence of the mean-field assumption of molecular chaos which predicts that the density fluctuations are equal to the ones of 
an ideal gas, e.g. 
are Poisson distributed. 
For example, using Eq. (\ref{I_ASYM}) we can rewrite the amplification coefficient $\Lambda$ from Eq. (\ref{LAMBDA_DEF}) as 
\begin{equation}
\label{NEW_LAM}
\Lambda={\sqrt{\pi} \over \eta}{\rm sin}\left({\eta\over 2}\right)
\big\langle (n+1)^{1/2}\big\rangle
\end{equation}
where the average of an arbitrary function $g(n)$ is defined as
\begin{eqnarray}
\nonumber
\langle g(n) \rangle &\equiv & \sum_{n=0}^{\infty} {\rm e}^{-M_R} {M_R^n\,g(n)\over n!} \\
\label{AVERAGE_DEF}
                             &=& \sum_{n=1}^{\infty} {\rm e}^{-M_R} {M_R^{n-1}\, g(n-1)\over (n-1)!} \\
\end{eqnarray}
Expanding $g$ around the mean value $M$ of the distribution 
one obtains,
\begin{equation}
\label{TAYLOR_G}
\langle g(n) \rangle =g(M)+\sum_{k=1}^{\infty}g^{(k)}(M) {\mu_k \over k!}
\end{equation}
where $g^{(k)}(M)$ is the $k-th$ derivative of $g$ evaluated at $n=M$, and $\mu_k$ are the so-called central moments, 
\begin{equation}
\label{CENTRAL_DEF}
\mu_k=\langle (n-M)^k\rangle 
\end{equation}
which are known for the Poisson distribution. For example, one has
\begin{eqnarray}
\nonumber
\mu_1&=&0 \\
\nonumber
\mu_2&=&\mu_3=M\\
\label{CENTRAL_POISS}
\mu_4&=&M+3M^2
\end{eqnarray}
Thus, the following approximation is obtained,
\begin{equation}
\label{APPROX_G}
\langle g(n)\rangle=g(M)+{M\over 2} g''(M)+{M\over 6}g'''(M)+{M+3M^2\over 4!} g^{(4)}+\ldots
\end{equation}
To evaluate transport coefficients, averages over polynomials in $n$ are needed. 
Therefore, we investigate a generic example, $g_0=(n+\epsilon)^{\alpha}$
with some power $\alpha$ and a positive constant $\epsilon$.
The approximation, Eq. (\ref{APPROX_G}), then gives for small $\epsilon\ll M$
\begin{equation}
\label{APPROX_POLYN}
\langle g_0 \rangle \approx M^{\alpha}
\left(1+{\alpha(\alpha-1)\over 2 M} +{\alpha(\alpha-1)(\alpha-2)\over 6M^2}+{\alpha(\alpha-1)(\alpha-2)(\alpha-3)[(1/M)+3]\over 24 M^2} 
\right)
\end{equation}
This approximation becomes accurate for $M\gg 1$. For very large $M$ it suffices to replace $\langle g\rangle$ by $g(M)$.
Thus, replacing $n$ by $M$ in Eq. (\ref{NEW_LAM}),
leads to the leading order approximation for the amplification factor $\Lambda$,
\begin{equation}
\label{NEW_LAM2}
\Lambda\approx {\sqrt{\pi} \over \eta}{\rm sin}\left({\eta\over 2}\right)
\,M^{1/2}\,.
\end{equation}
Thus, fulfilling the critical noise condition $\Lambda=1$ 
in the limit of infinite density $M\rightarrow \infty$ requires that the critical noise $\eta_C$ becomes equal to $2\pi$.
Expanding $\eta_C$ around its infinite density limit gives, 
\begin{equation}
\label{LARGE_M_ETAC}
\eta_C\approx 2\pi -4 \sqrt{\pi\over M}\,\;\;\;{\rm for}\,M\gg 1
\end{equation}
Approximation (\ref{LARGE_M_ETAC}) leads to
\begin{eqnarray}
\nonumber
{\rm sin}\left({\eta_C\over 2}\right)&\approx & 2\sqrt{\pi\over M} \\
\label{SIN_APPROX}
{\rm sin}\left(\eta_C\right)&\approx & -4\sqrt{\pi\over M} 
\end{eqnarray}
Introducing the relative noise distance $\delta$ to the threshold,
\begin{equation}
\label{DELTA_DEF}
\delta={\eta_C-\eta\over \eta_C}
\end{equation}
we consider the difference of
the amplification factor to its value at the threshold, $\Lambda(\eta_C)=1$ and find,
\begin{equation}
\label{LAMBDA_EXPAND}
\Lambda-1\approx {\sqrt{\pi}\over 2\pi}\left[{\rm sin}\left({\eta\over 2}\right)-{\rm sin}\left({\eta_C\over 2}\right] \right) M^{1/2}\approx \delta {\sqrt{ \pi M}\over 2},\;\;\;{\rm for}\,\delta\ll 1\,.
\end{equation}
As a preliminary to calculate transport coefficients, we express the variables $p$, $q$, $S$ and $\Gamma$ from Eqs. (\ref{DEF_PQ}),
in terms of averages over the Poisson distribution,
\begin{eqnarray}
\nonumber
p&=&
{4\over \eta} \sin{(\eta)}
\langle (n+1) J_1(n+1)\rangle \approx 
{1\over 2 \eta} \sin{(\eta)}
\\
\nonumber
q&=&
{4\pi \gamma^2\over \eta} \sin{(\eta)}
\langle (n+2) J_2(n+2)\rangle \approx 
{\pi \gamma^2\over 2\eta} \sin{(\eta)}
\\
\label{EVAL_PQ}
S&=&
{8\pi\gamma^2\over \eta }\sin{\eta\over 2}
\langle (n+2) J_3(n+2)\rangle \approx
-{\pi^{3/2}\gamma^2\over 4\eta \sqrt{M} }\sin{\eta\over 2}
\\
\nonumber
\Gamma&=&
{8\pi^2\gamma^4\over 3 \eta}\sin{\eta\over 2}
\langle (n+3) J_4(n+3)\rangle \approx
-{\pi^{5/2}\gamma^4\over 4\eta \sqrt{M} }\sin{\eta\over 2}
\end{eqnarray}
and evaluate the averages to leading order by means of Eqs. (\ref{J_DEFS}).
Finally, using Eq. (\ref{SIN_APPROX}), at the threshold one obtains
\begin{eqnarray}
\nonumber
p(\eta_C)&\approx&
-{1\over \sqrt{\pi M}}
\\
\nonumber
q(\eta_C)&\approx&
-\gamma^2\sqrt{\pi \over M}
\\
\label{THRESH_PQ}
S(\eta_C)&\approx&
-{\pi \gamma^2\over 4 M}
\\
\nonumber
\Gamma(\eta_C)&\approx &
-{\pi^2 \gamma^4\over 4 M}
\end{eqnarray}
These expressions, Eqs. (\ref{THRESH_PQ}) are used to evaluate the transport coefficients from Table \ref{TAB2} in the limit of large  $M$ and at the flocking threshold.
Only the leading order contributions in the limit of large $M$ are given. 
For example, the product $Sq\sim M^{-3/2}$ in coefficients such as 
$q_4$, $k_3$ and so on is neglected because $\Gamma\sim M^{-1}$ decays less strongly in the infinite density limit.
The results are given in Table \ref{TAB3}.
\begin{table}
\begin{center}
{\renewcommand{\arraystretch}{1.4}
\large
\begin{tabular}{|r||c|c|c|}
\hline
$j$ & $h_j$      &    $q_j$      &      $k_j$     \\
\hline
\hline
$1$ & $-{1\over 8} $         &  $-{S\over 2}\approx {\pi \gamma^2\over 8 M} $     &  $ -{S\over 8 }\approx {\pi \gamma^2\over 32 M} $ \\
\hline
$2$ & $-{1+10p\over 96}\approx -{1\over 96} $    & $-{S\over 4}\approx {\pi \gamma^2\over 16 M}$   & $-{5\, S \over 96}\approx {5 \pi \gamma^2\over 384 M} $ \\
\hline
$3$ & ${q\over 2}\approx -{\gamma^2\over 2} \sqrt{\pi\over M} $          & $\Gamma+Sq \approx -{\pi^2 \gamma^4\over 4 M}$ & 
${\Gamma\over 4}+{Sq\over 4} \approx -{\pi^2 \gamma^4\over 16 M}$ \\
\hline
$4$ & ${q\over 4}\approx -{\gamma^2\over 4}\sqrt{\pi\over M}$           & ${\Gamma\over 2}+{3 \,Sq\over 2}\approx -{\pi^2 \gamma^4\over 8 M}$ & ${\Gamma\over 12}+{Sq\over 3} \approx -{\pi^2 \gamma^4\over 48 M}$ \\
\hline
$5$ & $ {q(1+10p)\over 48}\approx -{\gamma^2\over 48}\sqrt{\pi\over M} $ & ${\Gamma\over 24}+{13 \,Sq\over 24}\approx -{\pi^2 \gamma^4\over 96 M} $ & $ { 5\,Sq\over 48} \approx {5 \pi^3 \gamma^6\over 768 M^2}$ \\
\hline
\end{tabular}
}
\caption{The transport coefficients $h_j$, $q_j$ and $k_j$,
defined in Eq. (\ref{TENSOR1}), evaluated at the threshold $\eta=\eta_C$, in the limit $M\gg 1$.
}
\label{TAB3}
\end{center}
\vspace*{-5ex}
\end{table}

It is instructive to evaluate the momentum density of a homogeneous, ordered system, close to the threshold, at $\delta\ll 1$,
According to Eqs. (\ref{W0_DEF}), (\ref{LAMBDA_EXPAND}), and Table \ref{TAB3}, we find
\begin{equation}
\label{W0_RES}
w_0=(2\delta)^{1/2} \pi^{-3/4} \gamma^{-2} M^{3/4}
\end{equation}
Even if the distance to the flocking threshold $\delta$, is tiny, the amplitude $w_0$ eventually diverges in the infinite density limit. 
It also has a strong dependence on the mean free path ratio $\gamma=R/\lambda$ and diverges for infinite mean free path $\lambda$.
This seemingly unphysical behavior
is just a consequence of the particular choice of dimensionless variables where all lengths are measured in units of the mean free path $\lambda$, and all times in units of the time step $\tau$.
The amplitude $w_0$ is a measure of global polar order but
it is not a convenient variable in the large density limit. 
Instead, I ``renormalize'' $w_0$ be relating it to the polar order parameter $\Omega$, which is defined
as the average speed of a particle $\langle|{\bf v}|\rangle$ divided by the individual particle speed $v_0$.
This way, $\Omega$ will always be between zero and one, and a small $\Omega\ll 1$ is expected near the order-disorder transition.
One has 
\begin{equation}
\Omega={<|{\bf v}|>\over v_0}={w_0\over \rho}
\end{equation}
where $\rho$ is the dimensionless particle density. 
It is related to the regular number density of a two-dimensional system, $\tilde{\rho}$ 
by $\rho=\tilde{\rho} \lambda^2$ 
Using the definition,
$M=\pi R^2 \tilde{\rho}=\pi R^2 \rho/\lambda^2$ we find $\rho=M/(\pi \gamma^2)$ and
an expression for $\Omega$ is obtained,
\begin{equation}
\label{OMEGA_W0_CON}
\Omega={\pi \gamma^2 \over M} w_0
\end{equation}
Inserting $w_0$ from Eq. (\ref{W0_RES}), the order parameter near threshold follows as
\begin{equation}
\label{OMEGA_VS_DELTA}
\Omega=2^{1/2} \delta^{1/2} \pi^{1/4} M^{-1/4}
\end{equation}
In contrast to $w_0$, the order parameter does not depend on particle speed. 
This is the expected result because, (i) we work in the mean-field approximation,  and (ii)
the collisional contribution to the transport coefficients was neglected earlier.

\subsection{Stability analysis}
\label{sec:stability}

A long wavelength instability has been reported in Vicsek-type models \cite{bertin_06,bertin_09,ihle_11,ihle_13}.
This instability is strongest for longitudinal sound-wave-like perturbations and exists in the ordered phase 
at a range of noise values $\eta_S\le \eta \le \eta_C$,
next to the flocking threshold.
In this chapter, I reanalyse this instability in the large $M$-limit 
for a
longitudinal perturbation with wave vector
$\vec{k}=k\hat{n}$. 
Without loss of generality, the collective motion is assumed to go into the x-direction, $\hat{n}=\hat{x}$.
As small perturbation of amplitude $\delta \rho$ is imposed on the homogeneous particle density $\rho_0$. A similar perturbation
$\delta w$ is added to the x-component of the homogeneous momentum density $w_{x,0}=w_0$, where $w_0$ is given in Eqs. (\ref{W0_DEF}) and (\ref{W0_RES}),
\begin{eqnarray}
\nonumber
\rho&=&\rho_0+\delta \rho\; {\rm e}^{ikx+\omega t} \\
\nonumber
w_x&=&w_0+\delta w\; {\rm e}^{ikx+\omega t} \\
\label{STAB_START}
w_y&=&0\,,
\end{eqnarray}
and $\omega$ is the complex growth rate of the perturbation.
From the continuity equation, Eq. (\ref{CONTIN}), a simple relation between $\delta \rho$ and $\delta w$ follows,
\begin{equation}
\label{RHO_W_REL}
\delta \rho=-{ik\over \omega} \delta w
\end{equation}
The hydrodynamic equation for the momentum density, Eq. (\ref{NAVIER}), in linear order of the perturbations leads to
\begin{eqnarray}
\label{NAVIER_STAB1}
& & (\omega-k^2 h_1 +2 ik w_0 h_3)\,\delta w
+(-ik^3 h_2+ik w_0^2 h_3'-k^2 w_0 h_4)\,\delta \rho= \\ 
\nonumber
& &(\Lambda-1+ik w_0 q_1+3 w_0^2 q_3)\,\delta w+ \\
\nonumber
& &(-ik b+w_0 \Lambda'
-k^2 w_0 q_2+ik w_0^2 q_4+w_0^3 q_3' +ik w_0^2 k_3)\,\delta \rho
\end{eqnarray}
Here, all transport coefficients are evaluated for a homogeneous system with constant density $\rho_0$.
The density derivatives of the coefficients $\Lambda$, $h_3$ and $q_3$ are also needed
and, for example, are given as 
\begin{equation}
\Lambda' \equiv \left. {d \Lambda\over d\rho}\right|_{\rho=\rho_0}=  \pi \gamma^2 {\partial \Lambda \over \partial M}\,. 
\end{equation}
Using Eq. (\ref{NEW_LAM2}), one finds 
\begin{equation}
\label{LAM_STRICH1}
{\partial\Lambda\over \partial M}={1\over 2 \eta} \sqrt{\pi\over M} {\rm sin}\left({\eta\over 2}\right)
\end{equation}
This expression is then evaluated at the threshold with the help of Eq. (\ref{SIN_APPROX}), leading to
\begin{equation}
\label{LAM_STRICH2}
\Lambda'(\eta_C)\approx {\pi \gamma^2\over 2 M}
\end{equation}
Similiarly, one obtains
\begin{equation}
\label{Q3_STRICH}
q_3'(\eta_C)=\left.{d q_3\over d \rho}\right|_{\eta=\eta_C}\approx
{\pi^3 \gamma^6\over 8 M^2}\,,\;{\rm and}\;\;h_3'(\eta_C)\approx 0\,
\end{equation}
Substituting Eq. (\ref{RHO_W_REL}) into (\ref{NAVIER_STAB1}), a quadratic equation for $\omega$  is obtained,
\begin{equation}
\label{QUADRAT}
\omega^2+\alpha \omega=\beta
\end{equation}
with complex coefficients $\beta=\beta_{R}+i \beta_I$, $\alpha=\alpha_R+i \alpha_I$.
The real and imaginary parts of $\alpha$ and $\beta$ are found as,
\begin{eqnarray}
\nonumber
\alpha_R&=&1-\Lambda-3 w_0^2 q_3 -h_1 k^2 = -2 w_0^2 q_3 -h_1 k^2\\
\nonumber
\alpha_I&=&k w_0(2 h_3 -q_1) \\
\nonumber
\beta_R&=& k^2[-b + w_0^2 (q_4+k_3-h_3')+k^2 h_2] \\
\beta_I&=& k w_0[ -\Lambda'-w_0^2 q_3' + k^2 ( q_2- h_4)]
\end{eqnarray}
where Eq. (\ref{W0_DEF}) has been used to eliminate $\Lambda$.
Eq. (\ref{QUADRAT}) has two solutions, $\omega_{\pm}$, where $\omega_{+}$
is the one with the larger real part. Solving for the real part, one finds,
\begin{equation}
\label{RADICAL}
Re(\omega_{+})=-{\alpha_R\over 2}+{(\Delta^2+\nu^2)^{1/4}\over \sqrt{2}}
\left(1+{\Delta\over \sqrt{\Delta^2+\nu^2}}\right)^{1/2}
\end{equation}
with the abbreviations
\begin{eqnarray}
\nonumber
\Delta & \equiv &{\alpha_R^2-\alpha_I^2\over 4}+\beta_R \\
\label{DELTA_NU_DEF}
\nu &\equiv & {\alpha_R \alpha_I\over 2}+\beta_I
\end{eqnarray}
Instead of directly analysing this complicated expression, a different approach is pursued: 
Splitting the growth rate into its real and imaginary part, $\omega=\omega_R+i \omega_I$,
Eq. (\ref{QUADRAT}) can be rewritten as two equations for $\omega_R$ and $\omega_I$,
\begin{eqnarray}
\nonumber
\omega_R^2-\omega_I^2+\alpha_R \omega_R -\alpha_I \omega_I&=&\beta_R \\
\label{TWO_EQ}
2 \omega_R \omega_I + \alpha_R \omega_I + \alpha_I\omega _R&=& \beta_I
\end{eqnarray}
The sign of the real part $\omega_R$ determines whether the system is linearly stable.
We therefore eliminate $\omega_I$ from Eq.(\ref{TWO_EQ}) and obtain a quartic equation with real coefficients for the real part of $\omega$,
\begin{equation}
\label{QUARTIC1}
c_4 \omega_R^4 +c_3 \omega_R^3 + c_2 \omega_R^2 + c_1 \omega_R = c_0
\end{equation}
with coefficients
\begin{eqnarray}
\nonumber
c_4&=&4 \\
\nonumber
c_3&=&8 \alpha_R \\
\nonumber
c_2&=& \alpha_I^2+5 \alpha_R^2-4 \beta_R \\
\nonumber
c_1&=& \alpha_R[\alpha_I^2+\alpha_R^2-4 \beta_R] \\
\label{C_COEFFS}
c_0&=& \beta_I^2+\alpha_I \alpha_R \beta_I + \beta_R \alpha_R^2
\end{eqnarray}
This equation can be solved exactly by radicals as given in Eq. (\ref{RADICAL}).
However, in the high density limit $M\gg 1$ and near threshold $\Omega \ll 1$, 
the terms in the quartic equation can differ by several orders of magnitude and the analytical expressions 
become difficult to evaluate.
Furthermore, our aim is to gain a better physical understanding and to obtain simple expressions for relevant quantities like
the maximum growth rate of a perturbation of the ordered state. 
Therefore, in the next section, instead of relying on 
the exact solution of the quartic equation, various scaling regimes of $\omega_R$ are identified by 
different Ansatzes to 
solve 
Eq. (\ref{QUARTIC1}) approximately. 

\subsection{Analysis of the dispersion relation}
\label{sec:dispers}

\subsubsection{The limit of vanishing wave number}
\label{sec:smallK}

To see whether there is an instability at all,
the limit of zero wave number $k\rightarrow 0$ is investigated
by inserting the ``hydrodynamic'' Ansatz 
\begin{equation}
\label{SMALL_K_ANSATZ}
\omega_R=d_1 k^2+d_2 k^4+O( k^6) 
\end{equation}
into Eq. (\ref{QUARTIC1}) and
expanding the coefficients from Eq. (\ref{C_COEFFS}) in powers of $k$,
\begin{eqnarray}
\nonumber
c_3&\equiv & c_{30}+c_{32} k^2 +O(k^4) \\
\nonumber
c_2&\equiv&c_{20}+c_{22} k^2 +O(k^4) \\
\nonumber
c_1&\equiv&c_{10}+c_{12} k^2 +c_{14} k^4 +O(k^6)\\
\label{C_IN_K}
c_0&\equiv&c_{02}k^2+c_{04}k^4+c_{06} k^6+O(k^8) \\
\end{eqnarray}
with new coefficients that do not depend on $k$, for example,
\begin{eqnarray}
\nonumber
c_{02}&=&w_0^2\left\{
(\Lambda'+w_0^2q_3')^2+2w_0^2 q_3(\Lambda'+w_0^2q_3') (2h_3-q_1)+4 w_0^2 q_3^2(-b+w_0^2[q_4+k_3-h_3'])\right\} \\
\nonumber
c_{04}&=&w_0^2\left\{2(\Lambda'+w_0^2q_3')(h_4-q_2)+(q_1-2h_3)[-h_1 (\Lambda'+w_0^2q_3')+2 w_0^2q_3(q_2-h_4)]\right. \\
\nonumber        
& &\left. +4q_3(w_0^2q_3 h_2-bh_1+w_0^2h_1[q_4+k_3-h_3'])\right\} \\
\nonumber
c_{10}&=&-8w_0^6 q_3^3\\
\label{NEW_C_DEF}
c_{12}&=&w_0^2 q_3\left\{ -2w_0^2(2 h_3-q_1)^2-12 w_0^2 q_3 h_1+8[-b+w_0^2(q_4+k_3-h_3')]\right\}
\end{eqnarray}
Collecting terms of order $k^2$ and $k^4$, we find
\begin{eqnarray}
\label{D1_EQ}
d_1&=&{c_{02}\over c_{10}} \\
\label{D2_EQ}
d_2&=&{1\over c_{10}^3}
\left[
c_{04} c_{10}^2-c_{02}^2c_{20}-c_{10}c_{12} c_{02}
\right]
\end{eqnarray}
To simplify the analysis we evaluate the dispersion in the large density limit $M\gg 1$
and replace $w_0$, using Eq. (\ref{OMEGA_W0_CON}) by $w_0=\Omega M/(\pi \gamma^2)$.
In order to trace the origin of stabilizing and destabilizing effects, mathematical ``markers'' are put on
the transport coefficients. This is done by taking the asymptotic expressions from Table \ref{TAB3} and multiplying them 
with dummy variables that 
become equal to one in the limit $M\rightarrow \infty$. For example, we have
\begin{eqnarray}
\nonumber
q_3&= & -{\bar{q}_3 f^2 \over 4 M} \\
\nonumber
h_3&=&-{\bar{h}_3 f \over 2 \sqrt{\pi M}} \\
\nonumber
\Lambda'&=&{\bar{\Lambda}' f\over 2 M} \\
\nonumber
q_3'&=&{\bar{q}_3' f^3\over 8 M^2}\\
\label{MARKERS}
b&=&{\bar{b}\over 2}
\end{eqnarray}
where $f\equiv \pi \gamma^2$ and the ``marker''-variables are given by $\bar{q}_3$, $\bar{h}_3$, $\bar{\Lambda}'$ and so on.
Inserting these ``marked'' coefficients into Eqs. (\ref{NEW_C_DEF}) and (\ref{D1_EQ}), one finds that for vanishing order, $\Omega\rightarrow 0$, the coefficient 
$d_1$ is always 
positive but becomes negative for a strongly ordered state, $\Omega\approx 1$.
To obtain quantitative results and to investigate the sign change of $d_1$, the dominant positive and negative terms must be balanced.
Balance is achieved by assuming that the order parameter $\Omega$ is of order $1/M$ or smaller.
Mathematically, this is imposed by introducing 
the scaled order parameter
$\bar{\Omega}\equiv \Omega M$
and assuming that $\bar{\Omega}$ is at most of order one.
Replacing $\Omega=\bar{\Omega}/M$ in the equations for the coefficients $c_j$ and $c_{ij}$, Eqs. (\ref{C_COEFFS}) and (\ref{NEW_C_DEF}), 
and neglecting terms of higher order in $1/M$, the coefficients $c_{ij}$ dramatically simplify to,
\begin{eqnarray}
\nonumber
c_{02}&\approx &{\bar{\Omega}^2 \over 8 M^2}[2 (\bar{\Lambda}')^2-\bar{b}\bar{q}_3^2 \bar{\Omega}^2] \\
\nonumber
c_{04}&\approx& -{\bar{\Omega}^2 \bar{q}_3 \bar{b} \bar{h}_1 \over 16 M} \\
\nonumber
c_{06}&\approx &-{\bar{b} \bar{h}_1^2\over 128} \\
\nonumber
c_{10}&\approx&{\bar{\Omega}^6 \bar{q}_3^3 \over 8 M^3} \\
\nonumber
c_{12}&\approx& {\bar{\Omega}^2 \bar{q}_3 \bar{b} \over M} \\
\nonumber
c_{14}&\approx & {\bar{b}\bar{h}_1\over 4} \\
\nonumber
c_{20}&\approx& {5 \bar{\Omega}^4 \bar{q}_3^2\over 4 M^2}\\
\nonumber           
c_{22}&\approx & 2 \bar{b} \\
\nonumber
c_{30}&\approx & {4 \bar{\Omega}^2\bar{q}_3\over M} \\
\label{CIJ_COEFFS}
c_{32}&\approx & \bar{h}_1
\end{eqnarray}
Note that the mean free path dependence through the factor $f=\pi \gamma^2$ dropped out 
exactly from these expressions at any density.
By means of Eqs. (\ref{D1_EQ}) and (\ref{D2_EQ}) the low wave number approximation, Eq. (\ref{SMALL_K_ANSATZ}),
can now be easily evaluated with the coefficients $d_1$ and $d_2$ found as,
\begin{eqnarray}
\label{D1_FOUND}
d_1&\approx & {MB\over \bar{\Omega}^4 \bar{q}_3^3} \\
\label{D2_FOUND}
d_2&\approx & -{M^3 B\over \bar{\Omega}^8 \bar{q}_3^5}\left( {10 B\over \bar{\Omega}^2 \bar{q}_3^2}+8\right)\;\;\;{\rm with} \\
\nonumber
B&\equiv &2 (\bar{\Lambda}')^2-\bar{b}\bar{q}_3^2 \bar{\Omega}^2 
\end{eqnarray}
While $\bar{b}$, $\bar{q}_3$ and $\bar{\Lambda}'$ are just marker variables that are equal to one, their appearance tells us that
the low wave number behavior $k\rightarrow 0$ is controlled by only four transport coefficients: $b$, $q_3$, $\Lambda'$ and $\Lambda$.
The latter coefficient enters implicitly because according to Eq. (\ref{W0_DEF}), it is involved in the
calculation of the order parameter $\bar{\Omega}$.
The most interesting prediction of Eq. (\ref{D1_FOUND}) is that there is a critical order parameter 
\begin{equation}
\label{OMEGA_C}
\bar{\Omega}_C={\sqrt{2} |\bar{\Lambda}'| \over \sqrt{\bar{b}} \bar{q}_3}=\sqrt{2}
\end{equation}
below which $d_1$ is positive.
Thus, very close to threshold, for $0<\Omega< \sqrt{2}/M$, we do have a long wave length instability but there is restabilization slightly further away from threshold,
at $\Omega\ge \sqrt{2}/M$. Therefore, the width of the instability window (in $\Omega$-space) scales as $1/M$.
Given that $\bar{\Omega}$ is at most of order one, one sees that inside the instability region, the growth rate $\omega$ increases rapidly with $M$ since $d_1\sim M$
and $d_2\sim M^3$. In the high density limit $M\rightarrow \infty$ this suggests that the point $k=0$ becomes singular, e.g. 
the value for $\omega_R$ at $k$ slightly above zero is
very different from $\omega(k=0)=0$. This behavior is supported by the direct solution of Eq. (\ref{QUARTIC1}).
Moreover, inside the instability window (where $\bar{\Omega}<\sqrt{2}$), the small wave number expansion is only valid 
for $k\ll k_1$ where $k_1$ is defined through the equality of the first two terms
in the expansion of $\omega_R$, $|d_1| k_1^2=|d_2| k_1^4$.
This gives,
\begin{equation}
\label{K1_DEF}
k_1=\left|{d_1\over d_2} \right|^{1/2}\,. 
\end{equation}
Evaluating $k_1$ near the threshold to collective motion, that is for $\bar{\Omega}\ll 1$, we have $B\approx 2 (\bar{\Lambda}')^2$ and $k_1$ follows as
\begin{equation}
\label{K1_CALC}
k_1\approx {\bar{\Omega}^3 \over M} {\bar{q}_3^2\over 2\sqrt{5} |\bar{\Lambda}'|}
={\bar{\Omega}^3 \over M} {1 \over 2\sqrt{5}}\,,
\end{equation}
which goes to zero in the limit $M\rightarrow \infty$ and also vanishes for $\bar{\Omega}\rightarrow 0$.
Direct solution of the quartic equation (\ref{QUARTIC1}) shows that the low $k$ expansion of $\omega_R$ is of very limited use because 
$k_1$ goes to zero too rapidly.
In particular, the shape of the dispersion relation near the onset of collective motion, 
which is characterized by the maximum (positive) growth rate $\omega_{R,max}$ and the highest unstable wave number $k_0$
(where $\omega_R(k_0)=0$)
cannot be determined using a Taylor-expansion of $\omega_R$. 
This is because $k_0$ will be shown to scale with a smaller power
of $\bar{\Omega}/M$. Therefore, in the large $M$ limit, 
$k_0$ is outside the range of validity of the Taylor series, that is $k_0\gg k_1$. 
Different techniques to find $k_0$ are presented in the next section.

Furthermore, Eq. (\ref{D1_FOUND}) also tells us that the mathematical 
reason for the instability is a nonzero density derivative $\Lambda'$ and that the sign of $\Lambda'$ does not matter.
One also sees from Eqs. (\ref{D1_FOUND}) and (\ref{D2_FOUND}) 
that it is the combination of the coefficients $b$ and $q_3$ that initiate restabilization. The coefficient $b$ describes the pressure term in the 
Navier-Stokes-like hydrodynamic equation (\ref{NAVIER}) and is equal to the square of the speed of sound, 
whereas $q_3$ controls the nonlinear (cubic) stabilization of the momentum density.

\subsubsection{The shape of the dispersion relation at large wavenumbers}
\label{sec:shape}
In the previous section we discussed that the Ansatz $\omega_R=d_1 k^2+d_2 k^4+\ldots$ only gives a condition at what 
values of the order parameter $\Omega$ an instability occurs but fails to predict the shape of
the dispersion relation inside the instability region.
In this section, different scaling analyses of the quartic equation (\ref{QUARTIC1}) are presented that are valid at larger
wave numbers $k$. In particular, $k$ is assumed to be 
larger than the limit value $k_1$ from Eq. (\ref{K1_CALC}), and finally large enough to describe
the value $k_0$ above which $\omega_R$ becomes negative again.
This value is of particular interest because through the length scale $L_{inst}=2\pi/k_0$ it gives 
us an crude idea above which system size $L^*$ the high density 
Vicsek model developes soliton-like density waves \cite{chate_04_08,ihle_13}.
The maximum value of the dispersion relation, $\omega_{R,max}$, is important to provide a
lower bound $2\pi/\omega_{R,max}$ 
on the time scale such waves can pile up. 

To calculate $k_0$ and $\omega_{R,max}$ the following scaling Ansatz is made,
\begin{eqnarray}
\label{SCAL1}
\omega_R&\equiv &{\hat{\omega} \over M} \\
\label{SCAL2}
k& \equiv & {\hat{k}\over \sqrt{M}} 
\end{eqnarray}
which will be justified {\em a posteriori}.
The rescaled wave number $\hat{k}$ and the rescaled growth rate $\hat{\omega}$ are assumed to be of order one.
Inserting these definitions into the quartic equation (\ref{QUARTIC1}) and only keeping terms in the lowest
order of the small parameter $1/M$, a simple quadratic equation emerges,
\begin{equation}
\label{QUADR1}
\hat{\omega}^2+\hat{\omega}\left[ {\bar{\Omega}^2 \bar{q}_3\over 2} +\hat{k}^2 {\bar{h}_1\over 8}\right]
={\bar{\Omega}^2 B \over 16 \bar{b}}-\hat{k}^2 {\bar{\Omega}^2 \bar{q}_3 \bar{h}_1 \over 32}
-\hat{k}^4 {\bar{h}_1^2\over 256}\,.
\end{equation}
Its solution, written in terms of the unscaled variables,  is
\begin{eqnarray}
\nonumber
\omega_R&\approx &{\bar{\Omega}\over M } {|\bar{\Lambda}'|\over \sqrt{8\bar{b}}}
-{\bar{\Omega}^2 \bar{q}_3\over 4 M}-{k^2\bar{h}_1\over 16}
\\
\label{QUADR1_SOL}
&= & {\bar{\Omega} \bar{q}_3 \over 4 M}\left[\bar{\Omega}_C-\bar{\Omega}\right]
-{k^2\bar{h}_1\over 16}
\end{eqnarray}
where we used the definition of the critical order parameter, Eq. (\ref{OMEGA_C}) in the second line to visualize 
that $\omega_R$ can only be positive if $0<\bar{\Omega}<\bar{\Omega}_C$.
This expression gives as a first estimate of the maximum value of $\omega_R$,
\begin{equation}
\label{OMEGA_MAX_1}
\omega_{R,max}\approx 
{\bar{\Omega} \over 4 M}\left[\sqrt{2}-\bar{\Omega}\right]
\end{equation}
and justifies the initial scaling choice, $\omega_R\sim 1/M$ of Eq. (\ref{SCAL1}). 
Very close to the threshold, at $\bar{\Omega}\ll 1$, one finds, 
\begin{equation}
\label{OMEGA_MAX_SIMP}
\omega_{R,max}\approx {\bar{\Omega}\over \sqrt{8} M}={\Omega\over \sqrt{8}}
\end{equation}
with $\bar{\Omega}=M\Omega$ and $\bar{\Omega}_C=\sqrt{2}$.
Clearly, expression (\ref{QUADR1_SOL}) ceases to be valid at very small $k$ because in 
contrast to Eq. (\ref{SMALL_K_ANSATZ}), it does not
predict $\omega_R(k=0)=0$. 
Comparing neglected terms in Eq. (\ref{QUARTIC1}) with the surviving ones in (\ref{QUADR1}) gives the condition
\begin{equation}
\label{K2_EXPR}
k\gg k_2={\bar{\Omega} \over M} {|\bar{\Lambda}'| \over 2 \bar{b}}={\Omega\over 2}\;\;\;{\rm for}\;\bar{\Omega}\ll 1 
\end{equation}
for approximation (\ref{QUADR1_SOL}) to be valid, assuming proximity to the threshold to ordered motion. 
Near the restabilization point where $\bar{\Omega}\approx \bar{\Omega}_C$ the wave number restriction changes to 
\begin{equation}
\label{NEAR_RESTAB1}
k\gg {|\bar{\Lambda}'|\over M}\sqrt{5\over 2 \bar{b}^3}
\end{equation}
which can be realized by noting that
\begin{equation}
\label{B_VS_D}
B=\bar{b}\bar{q}_3^2(\bar{\Omega}_C^2-\bar{\Omega}^2)\approx 2 \bar{\Omega}_C \bar{b}\bar{q}_3^2(\bar{\Omega}_C-\bar{\Omega})
\end{equation}
and assuming $(\bar{\Omega}_C-\bar{\Omega})\ll 1$ when analyzing the terms in Eq. (\ref{QUARTIC1}).

Finally, setting $\omega_R=0$ in Eq. (\ref{QUADR1_SOL}) leads to an estimate for $k_0$, 
the wave number which delimitates the domain of unstable modes,
\begin{equation}
\label{K0_EXPR}
k_0\approx \sqrt{4 \bar{\Omega} \bar{q}_3[\bar{\Omega}_C-\bar{\Omega}] \over M \bar{h}_1}
=2 \sqrt{4 \bar{\Omega} [\sqrt{2}-\bar{\Omega}] \over M }
\end{equation}
which for $\bar{\Omega}\ll 1$ gives,
\begin{equation}
\label{K0_EXPR_CLOSE}
k_0\approx 2^{5/4} \sqrt{\bar{\Omega}\over M} {|\bar{\Lambda}'|^{1/2} \over \bar{h}_1^{1/2} \bar{b}^{1/4}}=
2^{5/4} \sqrt{\bar{\Omega}\over M}\,.
\end{equation}
The result $k_0\sim 1/\sqrt{M}$ confirms the scaling Ansatz, Eq. (\ref{SCAL2}). 
It is also in the range of validity, $k_0\gg k_2$, of the approximation (\ref{QUADR1_SOL})
because $\bar{\Omega}/M$ is 
always much smaller than $\sqrt{\bar{\Omega}/M}$ for $M\gg 1$ and $\bar{\Omega}<\sqrt{2}$.

Maximizing $k_0$ and $\omega_R(k_0)$ at fixed $M$ but variable $\bar{\Omega}$ leads to the ``most dangerous'' value
$\bar{\Omega}_D=\bar{\Omega}_C/2={1\over \sqrt{2}}$. 
According to Eq. (\ref{OMEGA_VS_DELTA}) this corresponds to a relative noise distance $\delta_D$,
\begin{equation}
\label{DELTA_D}
\delta_D=\delta(\bar{\Omega}_D)={1\over 4 \sqrt{4 M^3}}
\end{equation}
from threshold. For example, at $M=5$, the instability is strongest at $\delta=\delta_D=0.013$.   
Furthermore, calculating 
\begin{eqnarray}
\nonumber
k_0(\Omega_D)&=& \sqrt{2\over M} \\
\label{DANGER}
\omega_{max}(\Omega_D)&=& {1\over 8 M} 
\end{eqnarray}
we get a lower bound for the system size $L^{*}$  above which the homeogeneous ordered state is unstable
and we can extract a typical time scale for the formation of this instability, $T^{*}$,
\begin{eqnarray}
\label{L_MAX}
L^{*}&\approx &\tau v_0 {2\pi\over k_0(\Omega_D)}=\tau v_0 \pi\sqrt{2 M} \\
\label{TIME_SCALE}
T^{*}&\approx & \tau {2\pi\over \omega_{max}(\Omega_D)}= 16 \pi M\,\tau
\end{eqnarray}
Here, we had to reinsert the time step $\tau$ and the particle speed $v_0$ because $k$ and $\omega$ are dimensionless quantities.
Thus we arrive at the prediction that the lower bound for $L^{*}$ increases proportional to $\sqrt{M}$.
Comparing the prediction for $L^{*}$ from Eq. (\ref{L_MAX}) to the insert in Fig. 1 of Ref. \cite{ihle_11}
very good agreement with the lower curve is found for $M\ge 2$. This figure was obtained directly from the full dispersion relation,
evaluated by Mathematica, and without making any large density approximations.
We also find that the number of iterations to see the instability (in a system that is significantly larger than $L^{*}$), is proportional
to $M$.

\subsubsection{The dispersion relation at intermediate wavenumbers}
\label{sec:intermed}

So far, we have found an approximation for $\omega_R$ given by Eq. (\ref{SMALL_K_ANSATZ}) for small $k\ll k_1$ and
another one, Eq. (\ref{QUADR1_SOL}), for large $k\gg k_2$. 
However, between $k_1$ and $k_2$ there is a large gap in wave number space.
The closer one is to the threshold value, the more this gap widens up since the ratio $k_2/k_1=\sqrt{5}/\bar{\Omega}^2$
diverges for $\bar{\Omega}\rightarrow 0$.
The previous approximations did not allow us to find the most unstable $k_{max}$.
One of the goals of this section is to find $k_{max}$ which is expected to be 
inside the unexplored gap in wavenumber space.

A direct evaluation of the quartic equation (\ref{QUARTIC1}) using Mathematica indicated that the cubic and the linear term
do not seem to be relevant in this intermediate range of wave numbers, $k_1\ll k \ll k_2$.
To put this observation on a more solid ground, the following scaling Ansatz is made
\begin{eqnarray}
\nonumber
\omega&=&{\hat{\omega}\over M^{1+\phi}} \\
\nonumber
k&=&{\hat{k}\over M^{1+\alpha}} \\
\bar{\Omega}&=&{\hat{\Omega}\over M^{\beta}}
\end{eqnarray}
with unknown positive exponents $\alpha$, $\beta$ and $\phi$ and inserted into Eq. (\ref{QUARTIC1}).
To obtain a meaningful balancing of terms in this equation and to prevent contradictions the coefficients must
fulfill the conditions,
\begin{equation}
\phi={1\over 2}(\alpha+\beta)\;\;{\rm and}\;\; {\alpha\over 3} < \beta < \alpha
\end{equation}
At the lowest order in $1/M$ one finds a balance between the quartic term $\sim \omega^4$ and the $O(\omega^0)$ term, leading to the approximation
\begin{equation}
\label{OM_SQRT1}
\omega_R\approx \sqrt{\bar{\Omega}\over M}\left({B\over 32}\right)^{1/4}\,k^{1/2}
\end{equation}
Near threshold, $\bar{\Omega}\ll 1$ and setting the dummy variables to one, this becomes
\begin{equation}
\label{OM_SQRT2}
\omega_R\approx {1\over 2} \sqrt{\bar{\Omega}\over M}\,k^{1/2}
={1\over 2} (\Omega\, k)^{1/2} 
\end{equation}
Away from the restabilization region, that is assuming $B=O(1)$ and $\bar{\Omega}<1$, the approximation (\ref{OM_SQRT1})
is valid in the following window:
\begin{equation}
\label{INTERMED_LIM1}
{\bar{\Omega}^3\over M} \sqrt{32\over B}\,\bar{q}_3^2\ll k \ll 
{\bar{\Omega}\over M} \sqrt{B\over 8}\, {1\over \bar{b}} 
\end{equation}
For $\bar{\Omega}\ll 1$ and setting the dummy variables to one this window becomes
\begin{equation}
\label{INTERMED_LIM2}
k_3\equiv {4 \bar{\Omega}^3\over M} \ll k \ll 
{\bar{\Omega}\over 2 M}\equiv k_4 
\end{equation}
This means the scaling regime $\omega_R\sim \sqrt{k}$ is only clearly visible
very close to the threshold where $\bar{\Omega}$ is smaller or equal to about $0.1$.
Note, that $k_4$ happens to be equal to $k_2$, and $k_3$ has the same scaling $\sim \bar{\Omega}^3/M$
than $k_1$.

The approximation for $\omega_R$ at intermediate wavenumber, Eq. (\ref{OM_SQRT2}), increases monotonically with $k$, whereas
the expression for high $k$, Eq. (\ref{QUADR1_SOL}), decreases monotonically.
However, the regions of validity for these approximations do not overlap.
Therefore, they should not be equated to obtain an estimate of $k_{max}$ where $\omega_R$ reaches its maximum.
To estimate $k_{max}$, we first construct another approximation whose region of validity extends above $k_4$.
This is done by not only balancing terms of order $\omega_R^4$ and $\omega_R^0$ in Eq. (\ref{QUARTIC1}), but also including 
a contribution proportional to $\omega_R^2$. This leads to a quadratic equation in $\omega_R^2$, 
\begin{equation}
\label{MATCH_START}
4\omega_R^4+2\bar{b}k^2\omega_R^2={\bar{\Omega}^2 B\over 8 M^2}\,k^2
\end{equation}
with the solution
\begin{equation}
\label{MATCH_SOL1}
\omega_R={1\over 2} \left\{
-\bar{b} k^2 +\sqrt{(\bar{b} k^2)^2+{\bar{\Omega}^2 B k^2\over 2M^2}
}
\right\}^{1/2}
\end{equation}
Compared to Eq. (\ref{INTERMED_LIM2}), the range of validity is extended to 
\begin{equation}
\label{INTERMED_LIM_NEW}
k_3 \ll k \ll 
4^{1/5}
\left({\bar{\Omega}\over M}\right)^{3/5}\equiv k_5
\;\;\;{\rm for} \bar{\Omega}\ll 1
\end{equation}
Now, we have an overlap with the large $k$ approximation since $k_2\sim \bar{\Omega}/M \ll k_5\sim (\bar{\Omega}/M)^{3/5}$.
To match the results in the overlap region, the square root in Eq. (\ref{MATCH_SOL1}) is expanded for large $k$ 
with the result:
\begin{equation}
\label{FMATCH1}
\omega_R^2\approx 
{|\bar{\Lambda}'|^2 \bar{\Omega}^2\over 8 M^2 \bar{b}}
-{|\bar{\Lambda}'|^4 \bar{\Omega}^4\over 32 M^4 \bar{b}^3 k^2}+O(1/k^4)
\;\;\;{\rm for}\; { \bar{\Omega}\over M} \sqrt{B\over 2} \ll k \ll k_5
\end{equation}
From the high $k$ expression (\ref{QUADR1_SOL}) one finds
\begin{equation}
\label{FMATCH2}
\omega_R^2\approx {|\bar{\Lambda}'|^2 \bar{\Omega}^2\over 8 M^2 \bar{b}}-{k^2 \bar{h}_1 |\bar{\Lambda}'|\bar{\Omega}\over 8 M \sqrt{8\bar{b}}}
\;\;\;{\rm for}\; k\gg k_2 
\end{equation}
where a term proportional to $k^4$ could be neglected for $k\approx k_{max}$.
The two expressions (\ref{FMATCH1}) and (\ref{FMATCH2}) becomes equal at the particular wavenumber $k_{match}$, given by
\begin{equation}
\label{KMATCH1}
k_{match}={1\over 2^{1/8}} \left({\bar{\Omega}\over M}\right)^{3/4}
\end{equation}
It seems plausible to assume that $k_{match}$ is a good estimate of the wavenumber $k_{max}$. 
This means $k_{max}$ is expected to be proportional to $(\bar{\Omega}/M)^{3/4}$. This is confirmed by a more accurate
calculation of $k_{max}$ and $\omega_{max}$ for $\bar{\Omega}\ll 1$:
\begin{eqnarray}
\label{KMAX_DET}
k_{max}&=&\left({2\over 9}\right)^{1/8} \left({\bar{\Omega}\over M}\right)^{3/4}\\
\label{WMAX_DET}
\omega_{max}&=& {1\over \sqrt{8}} {\bar{\Omega}\over M}-{1\over 3^{1/2} 2^{7/4}}{\bar{\Omega}\over M}^{3/2}
\end{eqnarray}
The prefactor in Eq. (\ref{KMAX_DET}) is about $10\%$ smaller than the one in Eq. (\ref{KMATCH1}) but the scaling is the same.

\subsubsection{Verification of scaling predictions}

\begin{figure}
\begin{center}
\vspace{0cm}
\includegraphics[width=3.2in,angle=0]{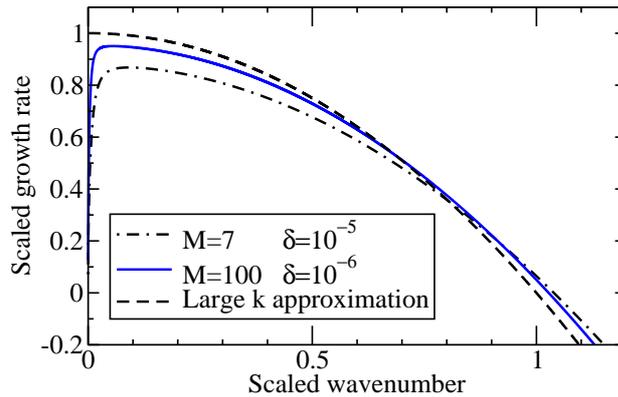}
\vspace{-1ex}
\caption{
The scaled growth rate $\omega_{R,S}=\omega_R\sqrt{8}/\Omega$ is plotted versus the scaled wavenumber $k_S=k\, 2^{-5/4}\Omega^{-1/2}$
for $M=7$ and $M=100$. $\omega_R$ was obtained from direct solution of Eq. (\ref{QUADRAT}).
The large $k$ approximation, Eq. (\ref{QUADR1_SOL}), is given by the dashed line and represents an inverted parabola.
The scaling was chosen to test the predictions for $k_0$ and $\omega_R$ from Eqs. (\ref{K0_EXPR_CLOSE}) and (\ref{OMEGA_MAX_SIMP}).
}
\label{FIG1}
\end{center}
\vspace*{-2ex}
\end{figure}
To verify the proposed scaling laws, Eq. (\ref{QUADRAT})
is evaluated numerically. 
Using Eq. (\ref{NEW_LAM})
the threshold value $\eta_C$ for a given $M$ is found from the condition $\Lambda(\eta_C)=1$.
For a given relative distance $\delta$, the noise is determined by $\eta=\eta_C(1-\delta)$.
The transport coefficients are obtained from Table \ref{TAB2}, by inserting
the large $M$ approximations for $p$, $q$, $\Gamma$ and $S$ from Eqs. (\ref{EVAL_PQ}), into the corresponding expressions.
Note that the approximations from Eq. (\ref{THRESH_PQ}) and from Table \ref{TAB3}
are not used.
Fig. \ref{FIG1} shows the scaled real part of the growth rate, $\omega_{R,S}=\omega_R\sqrt{8}/\Omega$, 
as a function of scaled wavenumber $k_S=k\, 2^{-5/4}\Omega^{-1/2}$ for $M=7$ and 
$\delta=10^{-5}$.
The values $k_{0}=0.15566$, $k_{max}=0.01372$ and 
$\omega_{R,max}=0.001233$ can be read off the plot.
Eqs. (\ref{OMEGA_W0_CON}) and (\ref{W0_DEF}) give the order parameters $\Omega=0.004018$ and $\bar{\Omega}=\Omega M=0.02813$ which are used
to calculate the limiting wavenumbers, $k_2$ to $k_5$. Following Eqs. (\ref{K2_EXPR}), (\ref{INTERMED_LIM2}) and 
(\ref{INTERMED_LIM_NEW}) one finds $k_2=k_4=0.02$, $k_3=1.27\times 10^{-5}$, and $k_5=0.0482$.
As expected, the conditions $k_0\gg k_2$, and $k_3\ll k_{max}\ll k_5$ are fulfilled.
The limit $k_1=7.1\times 10^{-7}$ is too small to play any role in the plot.
For verification, Eqs. (\ref{K0_EXPR}), (\ref{KMAX_DET}) and (\ref{WMAX_DET}) are evaluated and give 
$k_{0}=0.1507$, $k_{max}=0.01322$ and 
$\omega_{R,max}=0.001377$.
These values show excellent agreement with the ones from Fig. \ref{FIG1}.

I also checked a denser system with $M=100$ and $\delta=10^{-6}$ corresponding to $\bar{\Omega}=0.05741$.
Again, the errors of the predictions for $k_0$, $k_{max}$ and $\omega_{max}$ are only between 2 and 5 percent.
Finally, a very dense system with $M=8000$, $\delta=10^{-9}$ and $\bar{\Omega}=0.304$ was tested.
Here, the error of the prediction for $k_0$ was only $0.6\%$.
Since the scaling exponent of $3/4$ in the prediction for $k_{max}$ seems unusual, I checked whether the exponents
$1/2$ and $1$ could also make sense. It turns out that they can be ruled 
out because for the $M=7$ sample, $k_{max}$ would change by a factor of $\Omega^{1/4}\approx 4^{\pm 1}$, and for the $M=100$ 
system $k_{max}$ would change by a factor of $\approx 6.5^{\pm 1}$.
These factors are in clear contradiction with 
the small errors of the $k_{max}\propto \Omega^{3/4}$ prediction.

\subsection{Discussion of the dispersion relation} 

Close to the threshold to collective motion, that is for $\bar{\Omega}=\Omega M\ll 1$, 
the dispersion relation has a range of unstable modes.
One of the main results of this paper can be seen in Fig. \ref{FIG1}:
For $M\gg 1$, the unstable modes are asymptotically described by an inverted parabola 
for the real part of the growth rate $\omega_R$. It is not a perfect parabola since 
close to zero wavenumber, $\omega_R$ abruptly turns down. 
Inside this region of rapid change, a power law applies, $\omega_R\sim k^{1/2}$.
The imperfection of the parabola becomes smaller and smaller with increasing density, as shown in Fig. \ref{FIG1}.
Loosely speaking, I found a ``boundary layer'' in the dispersion relation.
For extremely small wavenumbers $k\approx 0$ there is another power law regime $\omega\sim k^2$. However,
for $\Omega M\ll 1$ and $M \gg 1$
this regime is basically irrelevant. For example, even for $M=7$ it is not visible in Fig. \ref{FIG1}.

Another main result is the scaling of the most unstable mode $k_{max}\sim \Omega^{3/4}$ and
$\omega_{R,max}\sim \Omega$, as well as the scaling of the range of unstable modes, $k_0\sim \Omega^{1/2}$.
The different scaling laws for $k_0$ and $k_{max}$, 
near the threshold to collective motion can be combined in the relation 
\begin{equation}
\label{SCALE_COMBIN}
{k_0^3\over k_{max}^2}=const.
\end{equation}
Note, that this disagrees with the low density Boltzmann approach for a Vicsek-like model by Bertin {\em et al} \cite{bertin_09}. 
The scaling reported there corresponds to $k_0^2/k_{max}=const.$

\section{Conclusion}

Previously, an Enskog-type kinetic theory for Vicsek-type models was proposed and 
hydrodynamic equations were drived \cite{ihle_11}.
The corresponding transport coefficients were given in terms
of infinite series which are difficult to handle. 
The fact that this theory is not restricted to low density has not been fully utilized yet.
In this paper, I exploit the special properties of the Poisson distribution to
obtain simple approximations of the transport coefficients. These expressions become exact
in the infinite density limit but, in practice, are expected to be 
quite good as long as $M$, the average number of collision partners, is of order one or larger.
Analyzing the hydrodynamic theory of the VM in the large $M$-limit 
is not only advantageous with respect to having simpler expressions but
also because the underlying mean-field assumption is supposed to become exact near the flocking threshold for $M\rightarrow \infty$.
Ranking transport coefficients by powers of the small parameter $1/M$ allows me to 
analytically evaluate the well-known density instability of the polarly ordered phase near the flocking threshold.

The growth rate $\omega$ of a longitudinal perturbation is calculated and a band of unstable modes with wavenumbers $0<k<k_0$ is found.
Some of the main results of this paper are given in Eqs. (\ref{K0_EXPR}), (\ref{KMAX_DET}) (\ref{WMAX_DET}) which describe 
the scaling behavior
of $k_0$ as well as the maximum growth rate and the most unstable mode number $k_{max}$ in terms of density and size of the order 
parameter $\Omega$.
It is found that there is only an instability for $0<\Omega<\sqrt{2}/M$. This means, for large $M$, 
a restabilization of the ordered phase occurs very close to the threshold -- the size of the instability window in $\Omega$-space shrinks as $1/M$.
Thus, one can assume that for large enough $M$, the restabilization can be described within the validity domain of hydrodynamic theory.
Inside the instability window, the largest value for $k_0$ is found for an order parameter value $\Omega=1/(M\sqrt{2})$.
This allows the calculation of the (approximate) 
maximum system size $L^{*}\sim \sqrt{M}$ below which the homogeneously ordered phase
is stable. A corresponding time scale for the formation of density inhomogeneities was also calculated and found to increase 
linearly with density.
The estimate for $L^{*}$ is in agreement with an earlier numerical evaluation of the dispersion relation
{\em without} the high density approximation, see Ref. \cite{ihle_11}.

Furthermore, I show that the real part of the growth rate follows three different power laws:
at very low wavenumber 
$k$, one has $Re(\omega)\sim k^2$ followed by a power law regime with $\sim \sqrt{k}$. At large enough wavenumber, one finds 
$Re(\omega)\sim const -k^2$ which represents an inverted parabola.
The closer the system is moved towards the threshold, 
the more room in $k$-space is taken by the inverted parabola, at the expense of the 
other two regimes.  
It is also interesting to see that the key features of the dispersion relation are determined by only five
($b$, $q_3$, $h_1$, $\Lambda$, and $|\Lambda'|$) of the many 
transport coefficients of the hydrodynamic equations.

Support
from the National Science Foundation under grant No.
DMR-0706017 
is gratefully acknowledged.


\begin{thebibliography}{99}
\bibitem{FOOT1}
In this context,
``enslaved'' means that the higher order coefficients are functionals of the lower Fourier coefficients $g_0,g_1,h_0, h_1$, and depend
only indirectly on space and time
through the temporal and spatial variations of these lower order coefficients.
\bibitem{FOOT2}
The critical noise $\eta_C$ goes to $2\pi$ for $M\rightarrow \infty$, \cite{ihle_11}.
$\eta=2\pi$ means that particle directions
are completely randomized in every iteration, and all memory of previous orientations
is
destroyed: particles just perform independent random walks.
Ring-kinetic theory \cite{chou_14} predicts that
for $\eta=2\pi$ the common prefactor of the amplification factor for connected correlation functions becomes zero.
Thus, even for $\eta$ slightly smaller than $2\pi$ correlations will be strongly damped.
\end{thebibliography}
\end{document}